\documentclass[12pt]{article}
\usepackage{amsmath}
\usepackage{amssymb}
\usepackage{amsfonts}
\newcommand{\ba}{\begin{array}{l}}
\newcommand{\ea}{\end{array}}
\newcommand{\beq}{\begin{equation}}
\newcommand{\eeq}{\end{equation}}
\newcommand{\bea}{\begin{eqnarray}}
\newcommand{\eea}{\end{eqnarray}}
\usepackage{epsfig}
\usepackage{graphicx}
\usepackage{color}

\definecolor{dyellow}{rgb}{1.,0.8,.0}
\definecolor{myblue}{rgb}{.1,.1,.7}
\definecolor{dcyan}{rgb}{.0,.6,.6}
\definecolor{dmagenta}{rgb}{0.6,0.0,0.6}
\definecolor{brown}{rgb}{0.6,0.2,0.}
\definecolor{darkblue}{rgb}{.0,.0,0.5}
\definecolor{darkred}{rgb}{0.75,0.0,0.0}
\definecolor{orange}{rgb}{1.,.6,.0}
\definecolor{dorange}{rgb}{0.8,.4,.0}
\definecolor{darkgreen}{rgb}{0.0,0.6,0.0}
\definecolor{purple}{rgb}{.4,.0,.4}

\def\bc{\begin{center}}
\def\ec{\end{center}}
\def\be{\begin{eqnarray}}
\def\ee{\end{eqnarray}}

\newcommand{\omits}[1]{}

\begin{document}
\begin{center}
{\Large \bf {  Toward a Gravitation Theory \\in Berwald--Finsler Space }}\\
  \vspace*{1cm}
Xin Li \footnote{lixn@mail.ihep.ac.cn} and  Zhe Chang
 \footnote{changz@mail.ihep.ac.cn}\\
\vspace*{0.2cm} {\sl Institute of High Energy Physics,
Chinese Academy of Sciences\\
P. O. Box 918(4), 100049 Beijing, China}\\

\bigskip

\end{center}
\vspace*{2.5cm}

%
\begin{abstract}
Finsler geometry is a natural and fundamental generalization of
Riemann geometry. The Finsler structure depends on both coordinates
and velocities. It is defined as a function on tangent bundle of a
manifold.  We use the Bianchi identities satisfied by Chern
curvature to set up a gravitation theory in Berwald-Finsler space.
The geometric part of the gravitational field equation is
nonsymmetric in general. This indicates that the local Lorentz
invariance is violated. Nontrivial solutions of the gravitational
field equation are presented.

\vspace{1cm}
\begin{flushleft}
\end{flushleft}
\end{abstract}

\newpage

\section{Introduction}
The possible violation of Lorentz invariance have been proposed
within several models of quantum gravity (QG) as well as the Very
Special Relativity (VSR)\cite{Cohen}. A succinct list of QG
includes: tensor VEVs originated from sting field
theory\cite{Kostelecky1}, cosmologically varying moduli
scenarios\cite{Damour}, spacetime foam models\cite{Amelino},
semiclassical spin--network calculations in Loop
QG\cite{Gambini,Alfaro}, noncommutative geometry
gravity\cite{Hayakawa,Mocioiu,Carroll,Anisimov} and  brane--world
scenarios\cite{Burgess}. A common feature of these phenomenological
studies on Planck scale physics is introducing of  modified
dispersion relations (MDR) for elementary particles.  Girelli {\em
et al.}\cite{Girelli} proposed a possible relation between MDR and
Finsler geometry.  Gibbons {\em et al.}\cite{Gibbons}  pointed out
that VSR is Finsler geometry. In the VSR, CPT symmetry is preserved.
VSR has radical consequences for neutrino mass mechanism.
Lepton-number conserving neutrino masses are VSR invariant. The mere
observation of ultra-high energy cosmic rays and analysis of
neutrino data give an upper bound of $10^{-25}$ on the Lorentz
violation\cite{neutrino}.

The above facts imply that new physics may connected with Finsler
geometry. In facts, in 1941 Randers\cite{Randers} published his work
on possible application of Finsler geometry in physics. Properties
of Randers space have been investigated exhaustively by both
mathematicians and physicists\cite{Yasuda}-\cite{Lixin}.

In a recent paper\cite{Kostelecky2}, Kostelecky studied the effect
of gravitation in the Lorentz- and CPT-violating Standard Model
Extension (SME). The incorporation of  Lorentz and CPT violation
into general relativity based on Riemann-Cartan geometry was
discussed. It provided dominant terms in the effective low-energy
action for the gravitational sector, thereby completing the
formulation of the leading-order terms in the SME with gravity. It
shows that a generalized geometric framework is helpful in
constructing a unification theory of gravity and electromagnetism,
weak and strong interaction.

Finsler geometry is a natural and fundamental generalization of
Riemann geometry. The Finsler structure depends on both coordinates
and velocities. It is defined as a mapping function from tangent
bundle of a manifold to $R^1$.  S. S. Chern\cite{Chern} proved that
there is a unique connection in the Finsler manifold that is torsion
free and almost $g$-compatibility. We use the Bianchi identities
satisfied by Chern curvature to set up a gravitation theory in
Berwald-Finsler space. The geometric part of the gravitational field
equation is nonsymmetric in general. This indicates that the local
Lorentz invariance is violated. Nontrivial solutions of the
gravitational field equation are presented.

This paper is organized as follows. In Sec. 2, we briefly review
basic concept and notations of Finsler geometry\cite{Book by Bao}.
The torsion free Chern connection and corresponding curvature are
introduced. The first and second Bianchi identities for curvature
are presented. Sec. 3 is devoted to construct a gravitation theory
in Berwald-Finsler space. Solutions of gravitational field equation
are shown in Sec. 4. In the final, we give conclusion and remarks.

\section{Finsler geometry}
\subsection{Finsler Manifold}
Denote by $T_xM$ the tangent space at $x\in M$, and by $TM$ the
tangent bundle of $M$. Each element of $TM$ has the form $(x, y)$,
where $x\in M$ and $y\in T_xM$. The natural projection $\pi :
TM\rightarrow M$ is given by $\pi(x, y)\equiv x$.

A Finsler structure of $M$ is a function\be F :
TM\rightarrow[0,\infty)\nonumber \ee with the following
properties:\\
(i) Regularity: F is $C^\infty$ on the entire slit tangent bundle
$TM\backslash0$.\\
(ii) Positive homogeneity : $F(x, \lambda y)=\lambda F(x,
y)$ for all $\lambda>0$.\\
(iii) Strong convexity: The $n\times n$ Hessian matrix\be
g_{ij}\equiv(\frac{1}{2}F^2)_{y^iy^j}\nonumber \ee is
positive-definite at every point of $TM\backslash0$, where we have
used the notation $(~)_{y^i}= \frac{\partial}{\partial y^i}(~)$.

Finsler geometry has its genesis in integrals of the form
\be\label{integral length} \int^r_s
F(x^1,\cdots,x^n;\frac{dx^1}{dt},\cdots,\frac{dx^n}{dt})dt.\ee

Throughout the paper, the lowering and raising of indices are
carried out by the fundamental tensor $g_{ij}$ defined above, and
its matrix inverse $g^{ij}$. Given a manifold $M$ and a Finsler
structure $F$ on $TM$, the pair $(M, F)$ is called as a Finsler
manifold. It is obvious that the Finsler structure $F$ is a function
of $(x^i,y^i)$. In the case of $F$ depending on $x^i$ only ,  the
Finsler manifold  reduces to Riemannian Manifold.

The symmetric  Cartan tensor can be defined as \be
A_{ijk}\equiv\frac{F}{2}\frac{\partial g_{ij}}{\partial
y^k}=\frac{F}{4}(F^2)_{y^iy^jy^k}~, \ee  Cartan tensor vanishes if
and only if $g_{ij}$ has no $y$-dependence.  So that Cartan tensor
is a measurement of deviation from Riemannian Manifold.

Using Euler's theorem on homogenous function, we can get useful
properties of the fundamental tensor $g_{ij}$ and Cartan tensor
$A_{ijk}$
 \be g_{ij} l^i=F_{y^j},\\g_{ij}
l^il^j=1,\\\label{fundametal tensor}y^i\frac{\partial
g_{ij}}{\partial y^k}=0,~~y^j\frac{\partial g_{ij}}{\partial
y^k}=0,~~y^k\frac{\partial g_{ij}}{\partial y^k}=0,\\
\label{cartan tensor}y^iA_{ijk}=y^jA_{ijk}=y^kA_{ijk}=0,\ee where
$l^i\equiv\frac{y^i}{F}$.

\subsection{Chern Connection}
The nonlinear connection $N^i_j$ on $TM\backslash0$ is defined as
\be
N^i_j\equiv\gamma^i_{jk}y^k-\frac{A^i_{jk}}{F}\gamma^k_{rs}y^ry^s,\ee
where $\gamma^i_{jk}$ is the formal Christoffel symbols of the
second kind \be\label{formal Christoffel symbols}
\gamma^i_{jk}\equiv \frac{g^{is}}{2}(\frac{\partial g_{sj}}{\partial
x^k}+\frac{\partial g_{sk}}{\partial x^j}-\frac{\partial
g_{jk}}{\partial x^s}).\ee The invariant connection under the
transform $y\longrightarrow \lambda y$ is of the form \be\label{N}
\frac{N^i_j}{F}\equiv\gamma^i_{jk}l^k-A^i_{jk}\gamma^k_{rs}l^rl^s.\ee

As usually,  we define the covariant derivatives
$\nabla\frac{\partial}{\partial x^i}$ and $\nabla dx^i$ as

\be\label{nabla1} \nabla\frac{\partial}{\partial
x^i}\equiv\omega^j_i\frac{\partial}{\partial x^j},\\
\label{nabla2}\nabla dx^i\equiv-\omega^i_jdx^j,\ee where
$\omega^i_j$ is the connection 1-forms. The operator $\nabla$ have
the same linear property with the covariant derivatives defined on
Riemannian manifold.

Here, we introduce the Chern connection that is torsion
freeness\be\label{torsion freeness} dx^j\wedge\omega^i_j=0\ee and
almost $g$-compatibility\be\label{almost g-compatibility}
dg_{ij}-g_{kj}\omega^k_i-g_{ik}\omega^k_j=2A_{ijs}\frac{\delta
y^s}{F}. \ee A theorem given by S.~S.~Chern \cite{Chern} guarantees
the uniqueness
of Chern connection.\\
Theorem (Chern): Let $(M,F)$ be a Finsler manifold. The pulled-back
bundle $\pi^\ast TM$ admits a unique linear connection, called the
Chern connection. Its connection forms are characterized by the
structural equations (\ref{torsion freeness}), (\ref{almost
g-compatibility}).\\
We ignore the proof of the theorem, just give some consequence of it
directly. Torsion freeness is equivalent to the absence of $dy^i$
terms in $\omega^i_j$; namely, \be\label{Chern connection1}
\omega^i_j=\Gamma^i_{jk}dx^k,\ee together with the symmetry \be
\Gamma^i_{jk}=\Gamma^i_{kj}.\ee And almost $g$-compatibility implies
that \be\label{Chern connection2}
\Gamma^i_{jk}=\frac{g^{is}}{2}\left(\frac{\delta g_{sj}}{\delta
x^k}+\frac{\delta g_{sk}}{\delta x^j}-\frac{\delta g_{jk}}{\delta
x^s}\right),\ee where \be \frac{\delta}{\delta
x^i}\equiv\frac{\partial}{\partial
x^i}-N^j_i\frac{\partial}{\partial x^j}.\ee The dual basis of
$\frac{\partial}{\partial y^i}$ is \be \delta y^i\equiv
dy^i+N^i_jdx^j.\ee  As before, we prefer to work with \be
\frac{\delta y^i}{F}=\frac{1}{F}(dy^i+N^i_jdx^j),\ee which is
invariant under rescaling of $y$.

We will work on two new natural local bases that are dual to
each other:\\
$\{\frac{\delta}{\delta x^i},F\frac{\partial}{\partial y^i}\}$ for
the
tangent bundle of $TM\backslash0$,\\
$\{dx^i,\frac{\delta y^i}{F}\}$ for the cotangent bundle of
$TM\backslash0$.

One can check that the transformation law of Chern connection on
Finsler manifold is the same with Riemannian connection on
Riemannian manifold. This fact is useful to guide us define the
covariant derivative of a tensor.

Let $V\equiv V_i^j\frac{\partial}{\partial x^j}\otimes dx^i$ be an
arbitrary smooth local section of $\pi^\ast TM\otimes\pi^\ast T^\ast
M$. The definition (\ref{nabla1}), (\ref{nabla2}) and property of
operator $\nabla$ imply that the covariant derivatives of $V$ is \be
\nabla V\equiv(\nabla V)_i^j\frac{\partial}{\partial x^j}\otimes
dx^i,\ee where \be\label{covariant derivative} (\nabla V)_i^j\equiv
dV_i+V_i^k\omega^j_k-V_k^j\omega^k_i.\ee $\nabla V$ is a 1-form on
$TM\backslash0$. Thus, it  can be expressed in terms of the natural
basis $\{dx^i,\frac{\delta y^i}{F}\}$,
 \be (\nabla
V)^j_i=V^j_{i~|s}dx^s+V^j_{i~;s}\frac{\delta y^s}{F}. \ee Using
relation between the Chern connection and the connection 1-forms
$\omega^i_j$ (\ref{Chern connection1}),  we obtain  the horizontal
covariant derivative $V^j_{i~|s}$\be V^j_{i~|s}=\frac{\delta
V^j_i}{\delta x^s}+V^k_i\Gamma^i_{jk}-V^j_k\Gamma^k_{is},\ee and the
vertical covariant derivative $V^j_{i~;s}$ \be
V^j_{i~;s}=F\frac{\partial V^j_i}{\partial y^s}.\ee The treatment
for tensor fields of higher rank is similar with the methods used on
Riemannian manifold. Here, we give results of  covariant derivatives
of the fundamental tensor $g$ and the norm 1 vector $l$:\be
g_{ij|s}=g^{ij}_{~|s}=0,\\
g_{ij;s}=2A_{ijs}~~~\rm{and}~~~g^{ij}_{~|s}=-2A^{ij}_{~~s},\\
\label{l|1}l^i_{~|s}=l_{i|s}=0,\\
l^i_{~;s}=\delta^i_s-l^il_s~~~\rm{and}~~~l_{i;s}=g_{is}-l_il_s.\ee

\subsection{Curvature}
The curvature 2-forms of Chern connection are \be\label{curvature
2-forms} \Omega^i_j\equiv d\omega^i_j-\omega^k_j\wedge\omega^i_k.\ee
The expression of $\Omega^i_j$   in terms of the natural
basis$\{dx^i,\frac{\delta y^i}{F}\}$ is of the form
 \be\label{omiga}
\Omega^i_j\equiv\frac{1}{2}R^{~i}_{j~kl}dx^k\wedge
dx^l+P^{~i}_{j~kl}dx^k\wedge\frac{\delta
y^l}{F}+\frac{1}{2}Q^{~i}_{j~kl}\frac{\delta
y^k}{F}\wedge\frac{\delta y^l}{F}, \ee where  $R$, $P$ and $Q$ are
the $hh$-,$hv$-,$vv$-curvature tensors of the Chern connection,
respectively. The following property is manifest
 \be\label{R1}
R^{~i}_{j~kl}=-R^{~i}_{j~lk},\\
Q^{~i}_{j~kl}=-Q^{~i}_{j~lk}. \ee We are now at the position to
demonstrate the  Bianchi identities for the curvature.

Exterior differential of the structural equation (\ref{torsion
freeness}) gives \be\label{torsion freeness1} dx^j\wedge
d\omega^i_j=0. \ee The combination of equations (\ref{torsion
freeness1}) and (\ref{torsion freeness}) shows that
 \be\label{34}
dx^j\wedge\Omega^i_j=0. \ee Substituting equation (\ref{34}) into
(\ref{omiga}), we get \be \frac{1}{2}R^{~i}_{j~kl}dx^j\wedge
dx^k\wedge dx^l+P^{~i}_{j~kl}dx^j\wedge dx^k\wedge\frac{\delta
y^l}{F}+\frac{1}{2}Q^{~i}_{j~kl}dx^j\wedge\frac{\delta
y^k}{F}\wedge\frac{\delta y^l}{F}=0. \ee The three terms on the left
side are completely independent. Thus, all of them should vanish.
This gives identities
\be\label{R2} R^{~i}_{j~kl}+R^{~i}_{k~lj}+R^{~i}_{l~jk}=0,\\
\label{P1} P^{~i}_{j~kl}=P^{~i}_{k~jl},\\
Q^{~i}_{j~kl}=0.\ee Then,  the curvature 2-forms can be simplified
as
\be\label{curvature expression}
\Omega^i_j\equiv\frac{1}{2}R^{~i}_{j~kl}dx^k\wedge
dx^l+P^{~i}_{j~kl}dx^k\wedge\frac{\delta y^l}{F}. \ee

Tedious but straightforward manipulation of exterior differential on
the structural equation (\ref{almost g-compatibility})  gives \be
\Omega_{ij}+\Omega_{ji}=-2(\nabla A)_{ijk}\wedge\frac{\delta
y^k}{F}-2A_{ijk}\left[d(\frac{\delta
y^k}{F})+\omega^k_l\wedge\frac{\delta y^l}{F}\right].\ee It can be
rewritten into \be\label{curvature}
\frac{1}{2}(R_{ijkl}+R_{jikl})dx^k\wedge
dx^l&+&(P_{ijkl}+P_{jikl})dx^k\wedge\frac{\delta y^l}{F}\nonumber\\
&=&-A_{iju}R^u_{~kl}dx^k\wedge
dx^l-2(A_{iju}P^u_{~kl}+A_{ijl|k})dx^k\wedge\frac{\delta
y^l}{F}\nonumber\\
&&+2(A_{ijk;l}-A_{ijk}l_l)\frac{\delta y^k}{F}\wedge\frac{\delta
y^l}{F}, \ee where we have used the abbreviations \be
R^i_{~kl}&\equiv& l^jR^{~i}_{j~kl}\\
P^i_{~kl}&\equiv& l^jP^{~i}_{j~kl}. \ee Equalization of  three
different types of terms at two sides of equation (\ref{curvature}
shows identities
\be\label{R3} R_{ijkl}+R_{jikl}&=&-2A_{iju}R^u_{~kl} ,\\
\label{P2} P_{ijkl}+P_{jikl}&=&-2(A_{iju}P^u_{~kl}+A_{ijl|k}),\\
\label{A1} A_{ijk;l}-A_{ijk}l_l&=&A_{ijl;k}-A_{ijl}l_k.\ee The
formula (\ref{R1}) and identities (\ref{R2}),(\ref{R3}) enable us
get  the fourth property of  $hh$-curvature,
 \be\label{R4}
R_{klji}-R_{jikl}=(B_{klji}-B_{jikl})+(B_{kilj}+B_{ljki})+(B_{iljk}-B_{jkil}),
\ee where, for convenient, we have used the notation $B_{ijkl}\equiv
-A_{iju} R^u_{kl}$.  On Riemannian manifold,  the Cartan tensor
vanish. This  means that  $B_{ijkl}=0$  on Riemannian manifold.  The
familiar   properties of Riemannian curvature
\be
\tilde{R}_{ijkl}+\tilde{R}_{ijlk}=0,\nonumber\\
\tilde{R}_{ijkl}+\tilde{R}_{kjli}+\tilde{R}_{ljik}=0,\nonumber\\
\tilde{R}_{ijkl}+\tilde{R}_{jikl}=0,\nonumber\\
\tilde{R}_{ijkl}-\tilde{R}_{klij}=0,\nonumber \ee  can be deduced
directly from the four properties of  $hh$-curvature (\ref{R1}),
(\ref{R2}), (\ref{R3}) and (\ref{R4}). Making use of the identity
(\ref{P2}) and equations (\ref{cartan tensor}), (\ref{l|1}), we may
get a  constituent relation for $P_{ijkl}$, \be\label{constitutive
relation}
P_{ijkl}=-(A_{ijk|l}-A_{jkl|i}+A_{kil|j})+A_{ij}^{~~u}\dot{A}_{ukl}-A_{jk}^{~~u}\dot{A}_{uil}+A_{ki}^{~~u}\dot{A}_{ujl},
\ee where \be \dot{A}_{ijk}\equiv A_{ijk|l}l^s.\ee Contracting
$P_{ijkl}$ with $l^i$ in equation (\ref{constitutive relation}),  we
obtain an  important relation \be P_{jkl}\equiv
l^iP_{ijkl}=-\dot{A}_{jkl}.\ee  The expression of $R$ and $P$ can be
got by  substituting the formula (\ref{curvature 2-forms}) into
(\ref{curvature expression}),
 \be R^{~i}_{j~kl}&=&\frac{\delta
\Gamma^i_{jl}}{\delta x^k}-\frac{\delta \Gamma^i_{jk}}{\delta
x^l}+\Gamma^i_{hk}\Gamma^h_{jl}-\Gamma^i_{hl}\Gamma^h_{jk},\\
P^{~i}_{j~kl}&=&-F\frac{\partial\Gamma^i_{jk}}{\partial y^l}.\ee
These are counterparts of the Riemannian curvature expessed in terms
of the Christoffel symbols $\tilde{\Gamma}^i_{jk}$
 \be \tilde{R}^{~i}_{j~kl}&=&\frac{\partial
\tilde{\Gamma}^i_{jl}}{\partial x^k}-\frac{\partial
\tilde{\Gamma}^i_{jk}}{\partial
x^l}+\tilde{\Gamma}^i_{hk}\tilde{\Gamma}^h_{jl}-\tilde{\Gamma}^i_{hl}\tilde{\Gamma}^h_{jk}.\ee
Before ending the section, we present  the second Bianchi identity.
Exterior differential of the Chern connection (\ref{curvature
2-forms}) gives \be\label{second bianchi1}
d\Omega^i_j-\omega^k_j\wedge\Omega^i_k+\omega^i_k\wedge\Omega^k_j=0.
\ee Substituting (\ref{curvature expression}) into the above
equation, we obtain \be \frac{1}{2}dR^{~i}_{j~kl}\wedge dx^k\wedge
dx^l&+&dP^{~i}_{j~kl}\wedge dx^k\wedge\frac{\delta
y^l}{F}-P^{~i}_{j~kl}dx^k\wedge d(\frac{\delta y^l}{F})\nonumber\\
&=&\frac{1}{2}R^{~i}_{r~kl}\omega^r_j\wedge dx^k\wedge
dx^l-\frac{1}{2}R^{~r}_{j~kl}\omega^i_r\wedge dx^k\wedge
dx^l\nonumber\\
&&+P^{~i}_{r~kl}\omega^r_j\wedge dx^k\wedge \frac{\delta
y^l}{F}-P^{~r}_{j~kl}\omega^i_r\wedge dx^k\wedge \frac{\delta
y^l}{F}.\ee  To evaluate $d(\frac{\delta y^l}{F})$, we first rewrite
$\frac{\delta y^l}{F}$ as \be \frac{\delta y^l}{F}
dl^l+\Gamma^l_{jk}l^kdx^j+\frac{d F}{F}l^l.\ee Then, one has
\be\label{d delta y} d(\frac{\delta
y^l}{F})&=&dl^j\wedge\omega^l_j+l^jd\omega^l_j+dl^l\wedge\frac{d
F}{F}\nonumber\\
&=&l^j\Omega^l_j+l^j\wedge\omega^k_j\wedge\omega^l_k+(\frac{\delta
y^j}{F}-\omega^j_kl^k-l^j\frac{d
F}{F})\wedge\omega^l_j+(\frac{\delta y^l}{F}-\omega^l_kl^k)\wedge
\frac{d F}{F}\nonumber\\
&=&l^j\Omega^l_j+\frac{\delta
y^j}{F}\wedge(\omega^l_j-l_j\frac{\delta y^l}{F}),\ee here we have
used the identity\be l_i\frac{\delta y^i}{F}=\frac{d F}{F}\ee to get
the third equal.\\
Substituting formula (\ref{d delta y}) into (\ref{second bianchi1})
and noticing the torsion freeness of the Chern connection,   we
obtain \be\label{second bianchi2} & &\frac{1}{2}\nabla
R^{~i}_{j~kl}\wedge dx^k\wedge dx^l+\nabla P^{~i}_{j~kl}\wedge
dx^k\wedge\frac{\delta
y^l}{F}\nonumber\\
& &~=P^{~i}_{j~kl}l^tdx^k\wedge(\frac{1}{2}R^{~l}_{t~rs}dx^r\wedge
dx^s+P^{~l}_{t~rs}dx^r\wedge\frac{\delta
y^s}{F})-P^{~i}_{j~kl}l_rdx^k\wedge\frac{\delta
y^r}{F}\wedge\frac{\delta y^l}{F}.\ee  In natural basis, we can
rewrite  equation (\ref{second bianchi2}) into the form \be
\frac{1}{2}(R^{~i}_{j~kl|t}-P^{~i}_{j~ku}R^u_{~lt})
dx^k\wedge dx^l\wedge dx^t&&\nonumber\\
+\frac{1}{2}(R^{~i}_{j~kl;t}-2P^{~i}_{j~kt|l}&+&2P^{~i}_{j~ku}\dot{A}^u_{~lt})
dx^k\wedge dx^l\wedge\frac{\delta y^t}{F}\nonumber\\
&+&(P^{~i}_{j~kl;t}-P^{~i}_{j~kl}l_t)dx^k\wedge\frac{\delta
y^l}{F}\wedge\frac{\delta y^t}{F}=0.\ee The three terms in  the left
side  are completely independent.  Then, we get the following
 identities
 \be\label{second bianchi R| P}
R^{~i}_{j~kl|t}+R^{~i}_{j~lt|k}+R^{~i}_{j~tk|l}&=&P^{~i}_{j~ku}R^u_{~lt}+P^{~i}_{j~lu}R^u_{~tk}+P^{~i}_{j~tu}R^u_{~kl},\\
R^{~i}_{j~kl;t}&=&P^{~i}_{j~kt|l}-P^{~i}_{j~lt|k}-(P^{~i}_{j~ku}\dot{A}^u_{~lt}-P^{~i}_{j~lu}\dot{A}^u_{~kt}),\\
P^{~i}_{j~kl;t}-P^{~i}_{j~kt;l}&=&P^{~i}_{j~kl}l_t-P^{~i}_{j~kt}l_l.\ee

\section{Gravitation theory in Berwald space}
Einstein proposed successfully his general relativity in Riemannian
space to describe gravity. It is interest to investigate the
behaviors of gravitation  in a more general Finsler spaces. Let us
briefly recall the setup way of  the Einstein field equation on
Riemannian manifold.  One starts from the second Bianchi identities
on Riemannian manifold \be\label{Bianchi on Riemann}
\tilde{R}^{~i}_{j~kl|t}+\tilde{R}^{~i}_{j~lt|k}+\tilde{R}^{~i}_{j~tk|l}=0.\ee
The metric-compatibility  \be
\tilde{g}_{ij|k}=0~~~~\mathrm{and}~~~~\tilde{g}^{ij}_{~~|k}=0,\ee
and contraction of (\ref{Bianchi on Riemann}) with $\tilde{g}^{jt}$
gives that \be
\tilde{R}^{ji}_{~~kl|j}+\tilde{R}^{i}_{~l|k}-\tilde{R}^{i}_{~k|l}=0,
\ee where $\tilde{R}^{i}_{~l}\equiv\tilde{R}^{ij}_{~jl}$ is the
Ricci tensor. Lowering the index $i$ and contracting with
$\tilde{g}^{ik}$, we obtain
 \be
\tilde{R}^{j}_{~l|j}+\tilde{R}^{j}_{~l|j}-\tilde{S}_{|l}=0 ,\ee
where $\tilde{S}=\tilde{g}^{ij}\tilde{R}_{ij}$ is the scalar
curvature. An equivalent but more familiar form is \be
(\tilde{R}^{jl}-\frac{1}{2}\tilde{g}^{jl}\tilde{S})_{|j}=0. \ee In
the weak field limit, gravitation theory should  reduce to the
Newtonian theory.   Einstein suggested his gravitational field
equation  of  the form \be
\tilde{R}_{jl}-\frac{1}{2}\tilde{g}_{jl}\tilde{S}=8\pi GT_{jl}, \ee
where $T_{jl}$ is the energy--momentum tensor and $G$ is the
Newton's constant.

In the paper, we use  similar approach to discuss gravitation  on
Finsler manifold.  Let us introduce first  two notions for Ricci
curvature: the Ricci scalar $Ric$ and the Ricci tensor $Ric_{ij}$.

The Ricci scalar is defined as \be Ric=g^{ik}R_{ik}, \ee where
$R_{ik}\equiv l^jR_{jikl}l^l$ is symmetric. The Ricci tensor on
Finsler manifold was first introduced by Akbar-Zadeh\cite{Akbar} \be
Ric_{ik}\equiv(\frac{1}{2}F^2Ric)_{y^iy^k},\ee which is manifestly
symmetric and covariant. Expanding  $y$ derivatives in the defining
formula for Ricci tensor $Ric_{ik}$, we get \be
Ric_{ik}=\frac{1}{4}(Ric_{;i;k}+Ric_{;k;i})+\frac{3}{4}(l_iRic_{;k}+l_kRic_{;i})+g_{ik}Ric
.\ee Substituting the defining formula for Ricci scalar $Ric$ into
the above equation, we obtain\be
Ric_{ik}&=&\frac{1}{2}(R^{~s}_{k~si}+R^{~s}_{i~sk})\nonumber\\
&&+\frac{1}{4}l^jl^l(R^{~s}_{j~sl;k;i}+R^{~s}_{j~sl;i;k})
-\frac{1}{4}l^jl^l(l_iR^{~s}_{j~sl;k}+l_kR^{~s}_{j~sl;i})\nonumber\\
&&+\frac{1}{2}l^j(R^{~s}_{i~sj;k}+R^{~s}_{j~si;k}+R^{~s}_{k~sj;i}+R^{~s}_{j~sk;i})\\
&=&\frac{1}{2}(R^{~s}_{k~si}+R^{~s}_{i~sk})+E_{ik}~,\ee where we
introduced the notation
 \be
E_{ik}&\equiv&\frac{1}{4}l^jl^l(R^{~s}_{j~sl;k;i}+R^{~s}_{j~sl;i;k})
-\frac{1}{4}l^jl^l(l_iR^{~s}_{j~sl;k}+l_kR^{~s}_{j~sl;i})\nonumber\\
&&+\frac{1}{2}l^j(R^{~s}_{i~sj;k}+R^{~s}_{j~si;k}+R^{~s}_{k~sj;i}+R^{~s}_{j~sk;i}).
\ee

Following same setup process for gravitational field equation in
Riemannian space, we start from the second Bianchi identities
(\ref{second bianchi R| P}). contracting it with $g^{jt}$, lowering
the index $i$, and contracting again with $g^{ik}$,  we get
\be\label{second bianchi deform1}
R^{ji}_{~~il|j}+R^{ji}_{~~lj|i}+R^{ji}_{~~ji|l}=g^{jt}g^{ik}(P_{jiku}R^{u}_{~lt}+P_{jilu}R^{u}_{~tk}+P_{jitu}R^{u}_{~kl}).\ee
Using the first Bianchi identity (\ref{R3}) and formula (\ref{R4}),
we can divide  the left side of the above equation into symmetric
part labeled by $[~]$ and  nonsymmetric part labeled by $\{~\}$ \be
R^{ji}_{~~il|j}&+&R^{ji}_{~~lj|i}+R^{ji}_{~~ji|l}\nonumber\\
&=&\left(Ric^j_{~l}+\frac{1}{2}B^{~kj}_{k~~l}-E^j_{~l}\right)_{|j}+\left(2B^{jk}_{~~lk}
+Ric^j_{~l}+\frac{1}{2}B^{~kj}_{k~~l}-E^j_{~l}\right)_{|j}-\delta^j_l(S-E)_{|j}\nonumber\\
&=&[(2Ric^j_{~l}-\delta^j_lS)-(2E^j_{~l}-\delta^j_lE)]_{|j}+\{B^{~kj}_{k~~l}+2B^{jk}_{~~lk}\}_{|j},
\ee where $E\equiv g^{ij}E_{ij}$ and $S=g^{ij}Ric_{ij}$. Using the
constituent relation of the $hv$-curvature tensor (\ref{constitutive
relation}), we rewrite the right side of identity (\ref{second
bianchi deform1})   as
\be
g^{jt}g^{ik}&(&P_{jiku}R^{u}_{~lt}+P_{jilu}R^{u}_{~tk}+P_{jitu}R^{u}_{~kl})\nonumber\\
&=&2(A^j_{~lu|i}-A^{jr}_{~~l}\dot{A}_{riu})R^{u~i}_{~j}+2(A^j_{~iu|j}-A_{u|i}+A^r\dot{A}_{riu}-A^{jr}_{~~i}\dot{A}_{rju})R^{u~i}_{~l},\ee
where $A_r\equiv g^{ij}A_{ijr}$.\\
Finally, we get the equivalent form of the identity (\ref{second
bianchi deform1}) \be\label{second bianchi deform2} &
&\left[\left(Ric^{jl} - \frac{1}{2}g^{jl}S\right)-
\left(E^{jl}-\frac{1}{2}g^{jl}E\right) \right]_{|j}
+\left\{\frac{1}{2}B^{~kjl}_k+B^{jkl}_{~~~k}\right\}_{|j}\nonumber\\
&=&(A^{jl}_{~~u|i}-A^{jrl}\dot{A}_{riu})R^{u~i}_{~j}+(A^j_{~iu|j}-A_{u|i}+A^r\dot{A}_{riu}-A^{jr}_{~~i}\dot{A}_{rju})R^{uli}.\ee
A Finsler structure $F$ is said to be of Berwald type if the Chern
connection coefficients $\Gamma^i_{jk}$ in natural coordinates have
no $y$ dependence. A direct proposition  on Berwald space is that
  $hv$--part of the Chern curvature vanishes identically
\be P^{~i}_{j~kl}=0, \ee and the $hh$--part of the Chern connection
reduce to \be R^{~i}_{j~kl}&=&\frac{\partial \Gamma^i_{jl}}{\partial
x^k}-\frac{\partial \Gamma^i_{jk}}{\partial
x^l}+\Gamma^i_{hk}\Gamma^h_{jl}-\Gamma^i_{hl}\Gamma^h_{jk}.\ee So
that, in Berwald space the identity (\ref{second bianchi deform2})
reduces as \be
\left[Ric^{jl}-\frac{1}{2}g^{jl}S\right]_{|j}+\left\{\frac{1}{2}B^{~kjl}_k+B^{jkl}_{~~~k}\right\}_{|j}=0
.\ee Thus, the counterpart of the Einstein's field equation on
 Berwald space  takes the form
 \be\label{field equation of Berwald}
\left[Ric_{jl}-\frac{1}{2}g_{jl}S\right]+\left\{\frac{1}{2}B^{~k}_{k~jl}+B^{~k}_{j~lk}\right\}=8\pi
G T_{jl}. \ee The gravitational field equation on Berwald space  is
obvious different from the Einstein's field equation.  The geometric
part contains nonsymmetric term. Thus,  in general,  the
energy--momentum tensor $T_{jl}$ is not symmetric.   It means that
local Lorentz invariance is violated in general.

\section{Solutions of gravitational field equation}
At this section, we present  examples of Berwald-Finsler space.
Kikuchi\cite{Kikuchi} proved that in a Randers space of Berwald
type, one  has \be\label{condition of Berwald} \tilde{b}_{i|j}\equiv
\tilde{b}_{i,j}-\tilde{b}_k\tilde{\gamma}^k_{ij}=0,\ee where
$\tilde{\gamma}^k_{ij}$ is the Christoffel symbols of Riemannian
metric $\tilde{a}\equiv \tilde{a}_{ij}dx^i\otimes dx^j$. In Randers
space,  one can derive straightforwardly the expression of the
geodesic spray coefficients as \be G^i\equiv
\gamma^i_{jk}y^jy^k=(\tilde{\gamma}^i_{jk}+l^i\tilde{b}_{j|k})y^jy^k+(\tilde{a}^{ij}-l^i\tilde{b}^j)(\tilde{b}_{j|k}-\tilde{b}_{k|j})\alpha
y^k,\ee and the Chern connection as\be
\Gamma^i_{jk}=(N^i_j)_{y^k}+\frac{1}{2}g^{it}y_s(N^s_t)_{y^jy^k}
.\ee It is not difficult to check that the geodesic spray
coefficients satisfy that \be\label{property of G}
\frac{1}{2}\frac{\partial G^i}{\partial y^j}=N^i_j.\ee Thus in
Randers spaces of Berwald type, the geodesic spray coefficients
reduce to \be G^i=\tilde{\gamma}^i_{jk}y^jy^k. \ee  The Chern
connection reduces to \be \Gamma^i_{jk}=\tilde{\gamma}^i_{jk}. \ee
Then, the $hh$--curvature  takes the form \be
R^{~i}_{j~kl}&=&\frac{\partial \tilde{\gamma}^i_{jl}}{\partial
x^k}-\frac{\partial \tilde{\gamma}^i_{jk}}{\partial
x^l}+\tilde{\gamma}^i_{hk}\tilde{\gamma}^h_{jl}-\tilde{\gamma}^i_{hl}\tilde{\gamma}^h_{jk}.
\ee In $4$-dimensional Randers space, the Robertson-Walker metric
\be \tilde{a}_{ij}={\rm
diag}\{1,-\frac{a^2(t)}{1-kr^2},-a^2(t)r^2,-a^2(t)r^2\sin^2\theta\}
\ee and the constraint \be \dot{a}^2+k=0 \ee gives nontrivial
solution of the gravitation in the Berwald-Finsler space.

A possible solution of (\ref{field equation of Berwald}) for
Berwald-Finsler space with one extra dimension is of the form
\be
\tilde{a}_{ij}&=&{\rm diag}\{1,-\frac{a^2(t)}{1-kr^2},-a^2(t)r^2,-a^2(t)r^2\sin^2\theta,0\},\\
\tilde{b}_i&=&\{0,0,0,0,c\}, \ee where $c$ is constant.

\section{Conclusion and  remarks}
In this paper, we have setup a gravitation theory in a torsion
freeness Berwald-Finsler space. The geometric part of the
gravitational field equation is , in general, nonsymmetric. This
fact indicates that the local Lorentz invariance is violated in the
Finsler manifold. This is in good agreement with discussions on
special relativity in Finsler space\cite{Gibbons, Girelli, Lixin}.
Nontrivial solutions of gravitation in Berwald-Finsler space were
presented.

However, problems  still remain. How to construct a gravitation in
general Finsler space is still a open question.  It is well-known
that in Riemannian space the sign of section curvature $K(x)$
determine the  type of geometry near $x$  (hyperbolic, flat or
spherical). In the landscape of Finslerian, the sign of $K(x,y)$
depend on the direction $y$ of our line of sight. This make it
possible to encounter all three types of geometry during a survey.
In such a cosmology model,   one may wish to find a natural
explanation for why the early universe is asymptotic flat.

\vspace{1.5cm}

\centerline{\bf \large Acknowledgements}

\vspace{0.3cm}

We would like to thank Prof. C.-G. Huang for helpful discussion. The
work was supported by the NSF of China under Grant No. 10575106.

\end{document}